\documentstyle [12pt,twoside]{article}

\def\be{\begin{equation}}
\def\te{\end{equation}}

\def\bea{\begin{eqnarray}}
\def\nn{\nonumber}
\def\tea{\end{eqnarray}}
\def\ba{\begin{displaymath}}
\def\ta{\end{displaymath}}

\def\utw#1{#1\llap{\lower2ex\hbox{$\widetilde{\hphantom{#1}}$}}}
\def\uttw#1{#1\llap{\lower2ex\hbox{$\widetilde{\widetilde{\hphantom{#1}}}$}}}

\def\f{\phi}

\def\frac#1#2{{\textstyle{#1\over#2}}}

\def\eg{{\it e.g.${~}$}}
\def\ie{{\it i.e.${~}$}}

\def\e{\epsilon}

\def\f{\phi}

\def\l{\lambda}

\def\s{\sigma}

\def\x{\xi}

\def\O{\Omega}

\def\Hat#1{\rlap{\kern.10em$\widehat{\phantom G}$}#1}
\def\HAt#1{\rlap{\kern.05em$\widehat{\phantom G}$}#1}

\def\cap#1{\rlap{\kern.1em$\widehat{\phantom{G\vrule height.8em}}$}#1{}}
\def\Cap#1{\rlap{\kern.05em$\widehat{\phantom{G\vrule height.8em}}$}#1{}}

\def\hg{{\hat{g}}}

\def\hA{{\Hat{A}}}
\def\hF{{\Hat{F}}}
\def\hN{{\Hat{N}}}
\def\hR{{\Hat{R}}}
\def\hs{{\hat{\s}}}

\def\utw#1{#1\llap{\lower2ex\hbox{$\widetilde{\hphantom{#1}}$}}}
\def\uttw#1{#1\llap{\lower2ex\hbox{$\widetilde{\widetilde{\hphantom{#1}}}$}}}
\def\uthN{\utw{\hN}}
\def\hD{\Hat{D}}
\def\hq{\hat{q}}
\def\hts{\widetilde{\hs}}

\def\begintitle#1#2#3#4#5#6
        {\begin{titlepage}
         \centerline{#1 \hfill CINVESTAV-GR-92-#2}
         \begin{center}\vglue .7in
         {\large\bf #3}\\
         {\large\bf #4}\\
         {\large\bf #5}\\[.7in]
         {\bf #6}\\[.3in]
         {\it Fisica - Centro de Investigaci\'on y de Estudios
         Avanzados, I.P.N.}\\
         {\it Apdo. Postal 14-740, 07000 Mexico D.F., Mexico}\\
         (e-mail: rcapovi@cinvesmx.bitnet)\\[.8in]
         {\bf ABSTRACT}\\
         \end{center}
         \begin{quotation}}

\def\endtitle
         {\end{quotation}
          \end{titlepage}
          \newpage}

\begin{document}

\begintitle{January 1992}{01}{NON-MINIMALLY COUPLED SCALAR FIELD}
{AND}{ASHTEKAR VARIABLES}{Riccardo Capovilla}
The non-minimal coupling of a scalar field is considered
in the framework of Ashtekar's new variables formulation
of gravity. A first order action functional
for this system is derived in which the field variables are
a tetrad field, and an SL(2,C) connection, together with
the scalar field. The tetrad field and the SL(2,C) connection
are related to the Ashtekar variables for the vacuum case by a conformal
transformation.
A canonical analysis shows that for this
coupling the equations of Ashtekar's
formulation of canonical gravity are non-polynomial in the
scalar field.

\vspace{1cm}


\endtitle

Ashtekar has shown that a convenient set of coordinates for
the phase space of general relativity is given by the
spatial self-dual spin connection, and the spatial triad,
its conjugate momentum \cite{Ashtekar1}.

The main advantage of this choice of coordinates is that
the equations of canonical vacuum general relativity take
a simpler form than in the conventional ADM coordinatization.
In particular, the equations turn out to be {\it polynomial}
in the field variables. The price to pay is that the use
of self-dual variables implies the use of complex variables,
for a space-time of lorentzian signature. Therefore one needs
to impose reality conditions a posteriori.
The price is not too high however. One can show that
the reality conditions can be put in polynomial form \cite{Ashtekar2}.

An important issue is to check if the key features of
this approach for vacuum general relativity survive
the coupling to matter fields. This question has been
addressed for matter field couplings of spin 1/2 and 3/2
by Jacobson \cite{J1,J2}, and systematically for
all physical matter field couplings by Ashtekar, Romano, and Tate
\cite{ART}.
The conclusion of these investigations is that in fact also in
presence of matter couplings the canonical equations
remain polynomial in the phase space variables, which
now include the matter fields.

In this note, we consider a formulation of the Ashtekar type of
the non-minimal coupling of a scalar field to gravity
via a term of the form $\x R \f^2$, where $\x$ is a dimensionless
constant, $R$ the curvature scalar, and $\f$ the scalar field.
Even if absent at the classical level,
in general a term of this form will be induced by the
quantization of an interacting scalar field in a curved
background (see \eg \cite{BPP}). To our knowledge, this
matter coupling has not been considered before in this
context. (The consequences of the presence of such a coupling
in the ADM framework have been studied recently
by Kiefer \cite{Kiefer}.)

We find that this matter coupling can be described by a first
order formalism which uses as field variables a tetrad field,
and an SL(2,C) connection, together with the scalar field.
The tetrad field and the SL(2,C) connection turn out to be
related to the analogous vacuum quantities by a scalar
field dependent conformal transformation. A canonical
analysis shows that this coupling gives field equations
that are {\it non-polynomial} in the field variables.
In particular, the
equations turn out to be non-polynomial in the scalar field.
The non-polinomiality is due to the presence of the
curvature scalar in the coupling.

The methods used in this paper are a natural generalization
to this particular coupling of the ones used in the vacuum
case, and for other matter couplings. Although we cannot
exclude the possibility of the existence of other strategies which may
give different results, no such alternative method
is apparent.

To conclude these introductory remarks, we note that our
analysis can be easily extended
to gravity theories of the Jordan-Brans-Dicke type, and to
the dilaton coupling, which emerges in the low energy limit
of string theory, reaching the same conclusions.
For concreteness, in the paper only the case of a non-minimally
coupled scalar field will be treated.

This paper is organized as follows. We start by deriving,
in two equivalent ways, a first order
Palatini-type action for a system composed of gravity and a non-minimally
coupled scalar field in the same spirit as the action for vacuum
general relativity of Refs. \cite{Sam,JS}.
We take as field variables
a tetrad field, and an SL(2,C) connection. Next, we
perform a canonical analysis of this action functional, and we
derive the hamiltonian for this system.

In our conventions we follow Ref. \cite{ART}. The space-time metric
has signature ($-+++$). Lower case latin letters denote
space-time (or spatial) indices. Upper case latin letters
denote SL(2,C) indices. They are raised and lowered using the
anti-symmetric symbol $\e_{AB}$, and its inverse, according
to the rules $\l^A = \e^{AB} \l_B $, $\l_A = \l^B \e_{BA}$.
We set $c = G = 1$.

\vspace{.5cm}

In the metric formalism the coupling
of interest is described by the action functional
\be
 S[g^{ab}, \f] = \int d^4 x \{ R \sqrt{-g} - 8\pi \sqrt{-g}\, [ \,
g^{ab} \partial_a \f \partial_b \f +(m^2 + \x R )\f^2 ] \},
\label{eq:act1}
\te
where $\x $ is a constant, and for the sake of simplicity we are
neglecting boundary terms.

We want to obtain a first order action functional
equivalent to (\ref{eq:act1}), in which the
field variables are a tetrad field, and an SL(2,C) connection.
The difficulty is given by the fact that we have
a derivative coupling. It is clearly not sufficient
just to consider the connection as an independent field variable.

Two strategies are natural. In the first, one uses the trick
of a scalar field dependent conformal transformation on the space-time
metric. This brings the action (\ref{eq:act1}) to a form in which there
is no derivative coupling.
The second strategy is analogous
to the one used to avoid the torsion that occurs in the Einstein-Cartan
theory of gravity.
The connection in (\ref{eq:act1}) is considered as a
field variable to be varied independently. The field equation
for the connection is then solved for the connection itself. This
yields an extra term in addition to the usual metric connection.
When
the solution is substituted back in the action one finds
an extra term that does not involve derivatives of the metric.
By subtracting this term from the initial action functional one obtains
the desired first-order action. This turns out to be equal
to the one obtained using the first strategy.
Since this fact does not seem evident a priori,
both strategies  will be described.

 In the first strategy,
one first rewrites the action (\ref{eq:act1})
in terms of the conformally transformed metric
\be
\hg_{ab} = \O^2 g_{ab},
\label{eq:cm}
\te
where the conformal factor is a function of the
scalar field defined by
\be
\O^2 := 1 - 8\pi \x \f^2.
\label{eq:cf}
\te
The action (\ref{eq:act1})  can then be written (up to a boundary term)
as
\be
 S [\hg^{ab}, \f ] = \int d^4 x \{ \hR \sqrt{-\hg}
- 8 \pi \sqrt{-\hg} [\Phi \hg^{ab} \partial_a \f
\partial_b \f + \O^{-4} m^2 \f^2 ]\},
\label{eq:act2}
\te
where we use a hat to denote quantities costructed from $\hg_{ab} $, and
we have defined the quantity
\be
\Phi := \O^{-2} [ 1 + 48 \pi \O^{- 2} \x^2 \f^2 ].
\label{eq:Phi}
\te
Note that we need to assume that the scalar field configuration
must be such that $\O^2 \ne 0$. This is necessary for (\ref{eq:act1})
and (\ref{eq:act2})  to yield
equivalent field equations. With this caveat, the use of the conformal
metric (\ref{eq:cm})  removes the difficulty
of a derivative  matter coupling.

It is now straightforward to write down a first order action
functional equivalent to (\ref{eq:act2}), since  (\ref{eq:act2})
is in the form of a minimally coupled scalar field (see \eg
Ref. \cite{JS}). We take as
field variables the tetrad field $\hs^a{}_{AA'}$, related to the
conformally transformed metric by $\hg^{ab} = \hs^a{}_{AA'} \hs^{bAA'}$,
and an SL(2,C) spin connection $\hA_a{}_A{}^B $,
with curvature
$\hF_{ab A}{}^B := 2 \partial_{[a} \hA_{b]A}{}^{B} +
2 \hA_{[a}{}_A{}^C \hA_{b]C}{}^B$.

In terms of these variables, an action functional equivalent to
(\ref{eq:act2}) is
\be
 S [\hs , \hA , \f ] = \int d^4 x \{ ({}^{(4)} \hs )
\hs^a{}_A{}^{A'} \hs^b{}_{BA'} \hF_{ab}{}^{AB}
+ 4\pi ({}^{(4)} \hs ) [\Phi \hg^{ab} \partial_a \f \partial_b \f +
\O^{-4} m^2 \f^2 ]\},
\label{eq:act3}
\te
where $({}^{(4)} \hs )$ denotes the determinant of $\hs_a{}^{AA'}$.
The $\hA_{aA}{}^B $ equation of motion identifies $\hA_{aA}{}^B$
as the self-dual part of the spin connection compatible with  $\hs^a_{AA'} $.
When this solution is substituted back in (\ref{eq:act3}), one
recovers (\ref{eq:act2}).

A different, and equivalent, strategy is to let
\begin{displaymath}
 \sqrt{-g} R (g_{ab}) \rightarrow
-2 ({}^{(4)} \s ) \s^a{}_A{}^{A'} \s^b{}_{BA'} \hF_{ab}{}^{AB}(\hA)
\end{displaymath}
in the action (\ref{eq:act1}). This gives
\be
 S [\s , \hA , \f ] = \int d^4 x \{ ({}^{(4)} \s ) \O^2
\s^a{}_A{}^{A'} \s^b{}_{BA'} \hF_{ab}{}^{AB}(\hA)
+ 4\pi ({}^{(4)} \s ) [ g^{ab} \partial_a \f \partial_b \f +
m^2 \f^2 ]\},
\label{eq:actin}
\te
where $\O$ is defined in (\ref{eq:cf}).
Varying this action
with respect to $\hA_{aA}{}^B $, and solving the resulting equation
for $\hA_{aA}{}^B $ gives
\be
\hA_a{}^{AB} = A_a{}^{AB} + \s_a{}^{(A}{}_{A'} \s^{|b|B)A'}
\partial_b \, ln \O,
\te
where $A_{aA}{}^B$ is the self-dual part of the spin connection
compatible with $\s_a{}^{AA'}$, with curvature
$R_{abA}{}^B$. When this solution is
substituted back in the action, one obtains, up to a boundary term,
\bea
 S [\s , \f ] &=& \int d^4 x \{ ({}^{(4)} \s ) \O^2
\s^a{}_A{}^{A'} \s^b{}_{BA'} R_{ab}{}^{AB}(\s) \nn \\
&+& 4\pi ({}^{(4)} \s ) [g^{ab} \partial_a \f \partial_b \f +
m^2 \f^2 ] - 3 (8 \pi \x \f \O^{-1})^2   ({}^{(4)} \s )
g^{ab} \partial_a \f \partial_b \f \}.
\label{eq:act5}
\tea
The last term in this action was not present in the initial second order
action (\ref{eq:act1}). What one can do is subtract it off from
the initial action (\ref{eq:actin}). This
results again in the action (\ref{eq:act3}).

\vspace{.5cm}

We turn now to the canonical analysis.
We consider a space-time that
is globally of the form $R \times \Sigma$. $\Sigma$ is a three dimensional
hypersurface. For simplicity $\Sigma$ is chosen to be  compact.

A $3+1$ decomposition
of the action (\ref{eq:act3}) yields
\bea
S &=& \int dt \int d^3 x [ i \sqrt{2} \hts^a{}_{AB}
\dot{\hA}_a{}^{AB}
+ i \sqrt{2}  \hA_0{}^{AB} \hD_a \hts^a{}_{AB} - i \sqrt{2}  \hN^a
\hts^b{}_{AB}
\hF_{ab}{}^{AB} \nn \\
&+& \uthN \hts^a{}^A{}_C \hts^b{}^C{}_B \hF_{ab}{}^B{}_A ]
+ 4 \pi \int dt \int d^3 x [\Phi \uthN \hts^a{}_{AB} \hts^b{}^{AB}
\partial_a \f \partial_b \f \nn \\
&-& \uthN^{-1} \Phi (\dot{\f} -\hN^a \partial_a \f )^2 + \uthN
\hts^2
\O^{-4} m^2 \f^2 ] .
\label{eq:act4}
\tea
Here $\hts^a_{AB} $ is the (densitized) spatial triad, \ie
$\hts^a_{AB} := (\hs ) \hs^a{}_{AB}$, with the spatial metric given by
$ \hq^{ab} = \hs^a{}_{AB} \hs^b{}^{AB} $, and $\hs $ the determinant
of $\hs_a{}^{AB}$.
$\uthN$ is the (densitized)
lapse function, $\uthN := (\hs )^{-1} \hN $, and $\hN^a$ the shift vector.
With an abuse of notation, $ \hA_a{}^{AB} $ denotes now the spatial part of
the SL(2,C) connection.  (Recall that hatted quantities refer to the
conformally
transformed metric (\ref{eq:cm}).)

The ``gravitational part" of (\ref{eq:act4})  is already in explicit
canonical form.
We need to reduce to canonical form the ``scalar field part".
The momentum conjugate to $\f$ is defined by
\be
 p := \partial L / \partial \dot{\f} =
- 8 \pi \Phi \uthN^{-1} (\dot{\f} - N^a \partial_a \f).
\te
With this expression, (\ref{eq:act4}) takes the form
\be
 S = \int dt \{ \int d^3 x [ i \sqrt{2} \hts^a{}_{AB}
\dot{\hA}_a{}^{AB} + p \dot{\f}] - H \},
\te
where
\bea
 H &=& \int d^3 x [- i \sqrt{2}  \hA_0{}^{AB} \hD_a \hts^a{}_{AB}
+ i \sqrt{2}  \hN^a\hts^b{}_{AB} \hF_{ab}{}^{AB} +
\uthN \hts^a{}^A{}_C \hts^b{}^C{}_B \hF_{ab}{}^B{}_A ] \nn \\
&-& \uthN ({p^2 \over 16\pi \Phi}
+ 4 \pi \Phi \hts^a{}_{AB} \hts^{b AB} \partial_a \f \partial _b \f
+ 4 \pi \hs^2 \O^{-4} m^2 \f^2 ) +  \hN^a p \partial_a \f ].
\tea
The configuration variables are $\hA_a{}^{AB}$, and the
scalar field $ \f $, with conjugate momenta
$\hts^a{}_{AB}$, and $p$, respectively. The variables
$\hA_0{}^{AB}, \uthN, \hN^a $ play the role of Lagrange multipliers,
enforcing the constraints corresponding to local gauge freedom,
and diffeomorphism invariance. The variables
$ \{ \hA_a{}^{AB} , \hts^a{}_{AB} \} $
are related to the Ashtekar's coordinates
for vacuum canonical gravity by a conformal transformation.

The hamiltonian is {\it not}  polynomial in the scalar field $\f$
because of the presence of the scalar field dependent function $\Phi$
in the denominator of the term involving $p^2$.
However, the hamiltonian is polynomial in the canonical pair
$ \{ \hA_a{}^{AB} , \hts^a{}_{AB} \} $.
As long as $\O \ne 0$, this implies that as for other matter couplings,
the phase space of this
system can be extended to include also degenerate metrics.

If additional non-derivative matter field couplings are present,
our analysis is extended by simply adding the appropriate
powers of the scalar field dependent $\O$,
wherever the space-time metric appears. For derivative
couplings, \eg spin 1/2, one would
need to redo the covariant analysis.
In both cases, additional matter fields would not modify our results.

The methods used in this paper
apply also to theories of the
Jordan-Brans-Dicke type, and to the dilaton coupling. For these systems,
it is always possible to find a conformal transformation of
the space-time metric which
brings the action to the form (\ref{eq:act2}), for {\it some}
field dependent function $\Phi$. The analysis parallels
the one given here, and the same type of non-polynomiality in the
scalar field
appears at the canonical level.

\vspace{.5cm}

\noindent ACKOWLEDGEMENTS

\vspace{.3cm}

I thank Ted Jacobson for suggesting this problem, and for his help
in the initial stages of this work.
This work was supported by a CONACyT post-doctoral fellowship,
and by the Centro de Investigaci\'on y de Estudios Avanzados del I.P.N.
(Mexico).

\end{document}